\RequirePackage{silence}
\documentclass{aa}

\usepackage{graphicx}
\usepackage{natbib}
\usepackage{scalerel}
\usepackage{comment}
\usepackage{placeins}
\usepackage[nolist,nohyperlinks]{acronym}
\usepackage[table]{xcolor}

\usepackage{newtxtext}
\usepackage{newtxmath}

\usepackage[utf8]{inputenc}
\usepackage[switch, modulo]{lineno}

\renewcommand{\linenumbers}[0]{}

\usepackage[dvipsnames,table]{xcolor}
\usepackage[normalem]{ulem}

\newcommand{\orcid}[1]{}

\begin{document}

\title{OGLE-2013-BLG-1125/MOA-2013-BLG-584: \\ Discovery of a gas giant orbiting an M dwarf}

\author{
E.~Bachelet\orcid{0000-0002-6578-5078}\thanks{\email{Corresponding author:etienne.bachelet@univ-fcomte.fr}}\inst{\ref{aff1}}
\and M.~Hundertmark\inst{\ref{aff2}}
\and V.~Bozza\inst{\ref{aff3},\ref{aff4}}
\and J.-P.~Beaulieu\orcid{0000-0003-0014-3354}\inst{\ref{aff5},\ref{aff6}}
\and A.~Udalski\inst{\ref{aff7}}
\and J.~C.~Yee\inst{\ref{aff8}}
\and Y.~Tsapras\inst{\ref{aff2}}
\and R.~A.~Street\inst{\ref{aff9}}
\and D.~M.~Bramich\inst{\ref{aff12}}
\and P.~Browne\inst{\ref{aff13}}
\and M.~Dominik\inst{\ref{aff13}}
\and K.~Horne\inst{\ref{aff13}}
\and S.~Ipatov\inst{\ref{aff14},\ref{aff15},\ref{aff16}}
\and C.~Snodgrass\inst{\ref{aff17}}
\and I.~A.~Steele\inst{\ref{aff18}}
\and R.~Figuera~Jaimes\inst{\ref{aff11},\ref{aff13}}
\and N.~Kains\inst{\ref{aff11}}
\and G.~D'Ago\inst{\ref{aff19}}
\and P.~Mróz\inst{\ref{aff7}}
\and J.~Skowron\inst{\ref{aff7}}
\and M.~K.~Szymański\inst{\ref{aff7}}
\and I.~Soszyński\inst{\ref{aff7}}
\and P.~Pietrukowicz\inst{\ref{aff7}}
\and S.~Kozłowski\inst{\ref{aff7}}
\and K.~Ulaczyk\inst{\ref{aff7},\ref{aff20}}
\and F.~Abe\inst{\ref{aff21}}
\and D.~P.~Bennett\inst{\ref{aff22},\ref{aff23}}
\and A.~Bhattacharya\inst{\ref{aff22},\ref{aff23}}
\and I.~A.~Bond\inst{\ref{aff24}}
\and R.~Hamada\inst{\ref{aff25}}
\and Y.~Hirao\inst{\ref{aff26}}
\and A.~Idei\inst{\ref{aff25}}
\and S.~Ishitani~Silva\inst{\ref{aff27},\ref{aff22}}
\and S.~Miyazaki\inst{\ref{aff28}}
\and Y.~Muraki\inst{\ref{aff21}}
\and T.~Nagai\inst{\ref{aff25}}
\and K.~Nunota\inst{\ref{aff25}}
\and G.~Olmschenk\inst{\ref{aff22}}
\and C.~Ranc\inst{\ref{aff5}}
\and N.~J.~Rattenbury\inst{\ref{aff29}}
\and Y.~K.~Satoh\inst{\ref{aff30}}
\and T.~Sumi\inst{\ref{aff25}}
\and D.~Suzuki\inst{\ref{aff25}}
\and T.~Tamaoki\inst{\ref{aff25}}
\and S.~K.~Terry\inst{\ref{aff22},\ref{aff23}}
\and P.~J.~Tristram\inst{\ref{aff31}}
\and A.~Vandorou\inst{\ref{aff22},\ref{aff23}}
\and H.~Yama\inst{\ref{aff25}}
\and M.~D.~Albrow\inst{\ref{aff32}}
\and D.~L.~DePoy\inst{\ref{aff33}}
\and A.~Gould\inst{\ref{aff34}}
\and C.-U.~Lee\inst{\ref{aff35}}
\and R.~W.~Pogge\inst{\ref{aff34},\ref{aff36}}
\and J.~W.~Blackman\inst{\ref{aff37}}
\and J.-B.~Marquette\inst{\ref{aff5}}
}
\institute{
Université Marie et Louis Pasteur, CNRS, Institut UTINAM UMR 6213, UFR Sciences \& Techniques, 16 route de Gray, 25030 Besançon Cedex, France\label{aff1}
\and Zentrum für Astronomie der Universität Heidelberg, Astronomisches Rechen-Institut, Mönchhofstr. 12--14, 69120 Heidelberg, Germany\label{aff2}
\and Università degli Studi di Salerno, Via Giovanni Paolo II, 132, 84084 Fisciano (SA), Italy\label{aff3}
\and Istituto Nazionale di Fisica Nucleare, Sezione di Napoli, Complesso Universitario di Monte Sant'Angelo, Edificio 6, Via Cintia, 80126 Napoli, Italy\label{aff4}
\and Institut d’Astrophysique de Paris, UMR 7095, CNRS, and Sorbonne Université, 98 bis boulevard Arago, 75014 Paris, France\label{aff5}
\and School of Natural Sciences, University of Tasmania, Private Bag 37 Hobart, Tasmania 7001, Australia\label{aff6}
\and Astronomical Observatory, University of Warsaw, Al. Ujazdowskie 4, 00-478 Warsaw, Poland\label{aff7}
\and Center for Astrophysics $|$ Harvard \& Smithsonian, 60 Garden
St., Cambridge, MA 02138, USA\label{aff8}
\and Las Cumbres Observatory, 6740 Cortona Drive, Suite 102, Goleta, CA 93117, USA\label{aff9}
\and European Southern Observatory, Karl-Schwarzschild-Str. 2, 85748 Garching bei München, Germany\label{aff11}
\and Center for Astrophysics and Space Science (CASS), New York University Abu Dhabi, P.O. Box 129188, Saadiyat Island, Abu Dhabi, United Arab Emirates\label{aff12}
\and SUPA School of Physics and Astronomy, University of St Andrews, North Haugh, St Andrews, Fife KY16 9SS, Scotland, United Kingdom\label{aff13}
\and Qatar Foundation, Education City, Doha, Qatar\label{aff14}
\and V.I. Vernadsky Institute of Geochemistry and Analytical Chemistry, Russian Academy of Sciences, 19 Kosygin Street, 119991 Moscow, Russia\label{aff15}
\and Space Research Institute of the Russian Academy of Sciences, Profsoyuznaya Str. 84/32, 117997 Moscow, Russia\label{aff16}
\and Max-Planck-Institut für Sonnensystemforschung, Justus-von-Liebig-Weg 3, 37077 Göttingen, Germany\label{aff17}
\and Liverpool John Moores University, Byrom Street, Liverpool L3 3AF, United Kingdom\label{aff18}
\and Institute of Astronomy, University of Cambridge, Madingley Road, Cambridge CB3 0HA, United Kingdom\label{aff19}
\and University of Warwick, Coventry CV4 7AL, United Kingdom\label{aff20}
\and Nagoya University, Furo-cho, Chikusa-ku, Nagoya, Aichi 464-8601, Japan\label{aff21}
\and NASA Goddard Space Flight Center, 8800 Greenbelt Road, Greenbelt, MD 20771, USA\label{aff22}
\and University of Maryland, College Park, MD 20742, USA\label{aff23}
\and Massey University, Tennent Drive, Palmerston North 4410, New Zealand\label{aff24}
\and Osaka University, 1-1 Yamadaoka, Suita, Osaka 565-0871, Japan\label{aff25}
\and The University of Tokyo, 7-3-1 Hongo, Bunkyo-ku, Tokyo 113-0033, Japan\label{aff26}
\and The Catholic University of America, 620 Michigan Ave.\ NE, Washington, DC 20064, USA\label{aff27}
\and Japan Aerospace Exploration Agency, Chofu Aerospace Center, 7-44-1 Jindaiji Higashi-machi, Chofu-shi, Tokyo 182-8522, Japan\label{aff28}
\and University of Auckland, 34 Princes Street, Auckland 1010, New Zealand\label{aff29}
\and Kanto Gakuin University, 1-50-1 Mutsuura-higashi, Kanazawa-ku, Yokohama 236-8501, Japan\label{aff30}
\and University of Canterbury, Ōtehīwai Mt John Observatory, 422 Godley Peaks Road, Lake Tekapo 7945, New Zealand\label{aff31}
\and University of Canterbury, Department of Physical and Chemical 
Sciences, Private Bag 4800, Christchurch 8020, New Zealand\label{aff32}
\and Department of Physics and Astronomy, Texas A\&M University, College Station, TX 77843-4242, USA\label{aff33}
\and Department of Astronomy, Ohio State University, 140 W.
18th Ave., Columbus, OH 43210, USA\label{aff34}
\and Korea Astronomy and Space Science Institute, Daejon
34055, Republic of Korea\label{aff35}
\and Center for Cosmology and AstroParticle Physics, Ohio State University, 191 West Woodruff Ave., Columbus, OH 43210, USA\label{aff36}
\and Physikalisches Institut, Universität Bern, Gesellschaftsstrasse 6, CH-3012 Bern, Switzerland
\label{aff37}}

\maketitle
\onecolumn

\parindent0pt
\vspace*{0.5em}
{\sf \em Context.}
{During the summer of 2013, survey and follow-up teams using conventional and robotic telescopes combined their efforts to identify and characterize planetary deviations in microlensing events towards the Galactic bulge.}

{\sf \em Aims.}
{We report the discovery and characterization of a planetary system in the microlensing event OGLE-2013-BLG-1125/MOA-2013-BLG-584.}

{\sf \em Methods.}
{We analyze photometric data from OGLE and MOA surveys, as well as LCO, $\mu$FUN, and Mindstep follow-up telescopes to model the microlensing light curve. We also used data from the Keck II telescope collected in 2018 to obtain additional constraints on the lensing system.}

{\sf \em Results.}
{We found a Saturn-mass planet of mass $M_{\rm p}=105\pm13 \ M_\oplus$ orbiting a low-mass host star $M_{\rm L}=0.139\pm0.011 \ M_\odot$ at $1.00\pm0.08$ kpc from Earth.}

{\sf \em Conclusions.}
{This discovery adds to the small sample of giant planets orbiting low-mass stars and provides constraints on planet formation theories.}

\parindent1.5em
\twocolumn
\section{Introduction}

 Gravitational microlensing is sensitive to planets with masses down to Earth-mass. This technique offers several advantages compared to other detection methods. First, it can detect host stars located several kiloparsecs away from Earth \citep{mao91, Gouldloeb1992} and therefore helps us to understand the planet and stellar formation throughout our Galaxy \citep{sno04,gau02,gou10,sum10,sum11,cas12,suz16,tsa16,ben96,Zhang2025}. Second, microlensing events arise from the chance alignment of a lens and a source star, implying that host stars are sampled approximately in proportion to their Galactic abundance. Consequently, M dwarfs are expected to constitute at least 50$\%$ of the host population \citep{Henry2006,Reyle2021}. These microlensing detections therefore contribute to our astrophysical understanding of ``planetary'' systems orbiting low-mass hosts \citep{miy18,jun18}, which potentially include brown dwarfs \citep{Hoy2026}. This is because gravitational microlensing does not require light from the host object to be detected. Indeed, observable light comes only from the more distant source star and is deflected when traversing the planetary system, which acts as the gravitational lens. Because the observed magnification depends on the exact alignment between lens, source, and observer, as well as the configuration of the lens itself, the mass ratio between the components of the lens and the projected orbital separation can lead to distinct observable features in an otherwise symmetric light curve.

Gravitational microlensing provides mass estimates by measuring the angular Einstein radius $\theta_{\mathrm{E}}$ and the microlensing parallax $\pi_{\rm E}$ \citep{gou92}:
\begin{equation}
M_{\rm L}=\frac{\theta_{\rm E}}{\kappa \pi_{\rm E}}, \kappa\equiv \frac{4 G}{c^2 \rm au} = 8.144 \ \frac{\rm mas}{M_\odot}.
\label{eq:mldl}
\end{equation}
Galactic gravitational microlensing images are separated on a scale of milli-arcseconds, and are thus typically not resolvable, except with the recent progress of interferometric observations \citep{Dong2019,Cassan2022,Mroz2025}. The relative proper motion between the lens and source changes the shape, size and location of these images giving rise to the observed lightcurve. $\theta_{\rm E}$ can be estimated if finite-source effects are observed in the lightcurve \citep{Yoo2004} or via the measurement of the lens-source relative proper motion $\mu_{\rm rel}$:
\begin{equation}
\mu_{\rm rel}=\frac{\theta_{\mathrm{E}}}{t_{\mathrm{E}}},
\end{equation}
where $t_{\rm E}$ is the Einstein ring crossing time. This can be achieved by resolving the lens from the source through high-resolution imaging observations conducted long after the event  \citep{Batista2015} or before the peak of the event \citep{Bachelet2022}. The planet mass is then derived as:
\begin{equation}
M_{p} =\frac{\mu_{\mathrm{rel}} t_{\mathrm{E}}}{\kappa \pi_{\mathrm{E}}} \frac{q}{1+q}, 
\end{equation}
where $q$ is the planet to host mass ratio. 

In this work, we present the analysis of the OGLE-2013-BLG-1125/MOA-2013-BLG-584 microlensing event. The photometric data shown in Section \ref{sec:discovery} exhibits the signature of a binary lensing event with an intermediate/resonant topology. { We present the high-resolution observations made with Keck in 2018 in Sect.~\ref{sec:KECK}, introduce the modeling procedure in Sect.~\ref{sec:analysis} and discuss the properties of the source in Section ~\ref{sec:sourceproperties}}. In Section ~\ref{sec:interpretation}, we present the methodology used to combine all the constraints to find that the lens system is a low-mass star orbited by a Saturn-mass planet. 

\section{Discovery and observations}
\label{sec:discovery}
\subsection{Photometric observations}

{On 2013 June 30}, the microlensing event OGLE-2013-BLG-1125 (J2000; $\alpha$:17:53:27.01, $\delta$:-29:47:34.1) was alerted by the Optical Gravitational Lensing (OGLE) team, carrying out the fourth phase of their experiment \citep{uda15}. The event was independently reported as MOA-2013-BLG-584 by the Microlensing Observations in Astrophysics collaboration (MOA; \citep{Bond2001}). They used a custom {$R$} filter (a combination of $V$ and $I$) for the majority of their observations. The MOA data used in this analysis were re-reduced to achieve optimal photometric precision, based on the work of Ian Bond \footnote{\url{http://www.massey.ac.nz/~iabond/staging/}}. Similarly, OGLE $I$ and $V$ data were re-processed with the OGLE custom pipeline, based on the work of \citet{Wozniak2000}. In addition, it was observed in the $K$ band by the VISTA variables in the Via Lactea (VVV) survey \citep{minniti2010}. The images were reprocessed to achieve superior photometric precision, as described by \citet{Beaulieu2018}.

In parallel, follow-up observations began on {2013 Sep 13} by the RoboNet team \citep{Burgdorf2007} thanks to their automated ranking method ({\tt RoboTAP})  \citep{hor09,dom12,hun17}, {which combines target} brightness, observability and planet sensitivity. The OGLE field number is BLG515, indicating that it is only observed once or twice every night, therefore boosting even more RoboTAP priority. Once the {event is selected}, observations were submitted on the southern telescopes of the Las Cumbres Observatory Global Telescope (LCO) network, namely the South African Astronomical Observatory (SAAO, but coded as CPT by LCO), the Siding Spring Observatory (SSO, coded as COJ by LCO) and the Cerro Tololo Interamerican Observatory (CTIO, LSC in LCOGT convention) sites. In addition, it was also requested by the ARTEMiS system \citep{dom08}. The Microlensing Network for the Detection of Small Terrestrial Exoplanets (MiNDSTEp) team started observing this event with their EMCCD camera \citep{sko15} on the Danish telescope {at La Silla (Chile)}, after an observing trigger was issued by the {\tt SIGNALMEN} anomaly detector \citep{dom07}, operated as part of the ARTEMiS system. The MicroFUN ($\mu$FUN) observations were triggered by an alert from T.\ Bensby at UT 15:51 on {September 22} to support a program for obtaining 8 m class spectra of highly magnified dwarfs and subgiants \citep{Bensby2017}. For OGLE-2013-BLG-1125, they obtained $T_{\rm eff}=4739\pm86$ K, $\log g =3.0\pm0.18$, ${\rm [Fe/H]} = -0.12\pm0.07$ for the source.

RoboNet and MiNDSTEp data were reduced using respective versions of the {\tt DanDIA} pipeline \citep{bra08}. Time-resolved color information is available from both the OGLE as well as the $\mu$FUN CTIO 1m telescope. Multi-color baseline observations for RoboNet were secured in 2016 when revisiting a number of observed fields in different filters. The VPHAS+ DR2 catalog also contains two entries for the event \citep{dre14}, which can be used for estimating the baseline measurements on a color-color diagram.

To speed up the modeling, we restricted the datasets to $2456310<{\rm JD}<2456965.25$. We verified that this cut did not impact the results in practice. We also noticed  a baseline variation several years after the peak, as detailed in Sect.~\ref{sec:KECK}. A summary of observations is shown in Table~\ref{table:sites}.

\subsection{Observations with Keck}
\label{sec:KECK}
We observed OGLE-2013-BLG-1125 with NIRC2 instrument on Keck II on {2018 May 25} in $K_S$ band centered at $\lambda_c=2.146$\, $~\mu$ m. At the time of observation, it could be expected that the source and lens should be separated by $\sim$40 mas, and therefore potentially observable with Keck. We used the Wide camera mode of NIRC2 with a pixel scale of 39.69 ${\rm mas.pixel^{-1}}$. We obtained 10 frames with 30 s exposure time, with a 5 position dither with an amplitude of 2 arcsec.
We processed the data using the {\tt KAI} reduction pipeline \citep{Lu2022} in a standard way, doing dark, flatfield corrections, masking bad pixels and cosmic rays before producing the stacked image. We obtained a 100 mas image. We performed aperture photometry using {\tt Sextractor} \cite{Bertin1996} from the {\tt AstrOmatic} package \cite{Bertin2012}.

\begin{figure}[t]
\centering
\includegraphics[width=0.45\columnwidth]{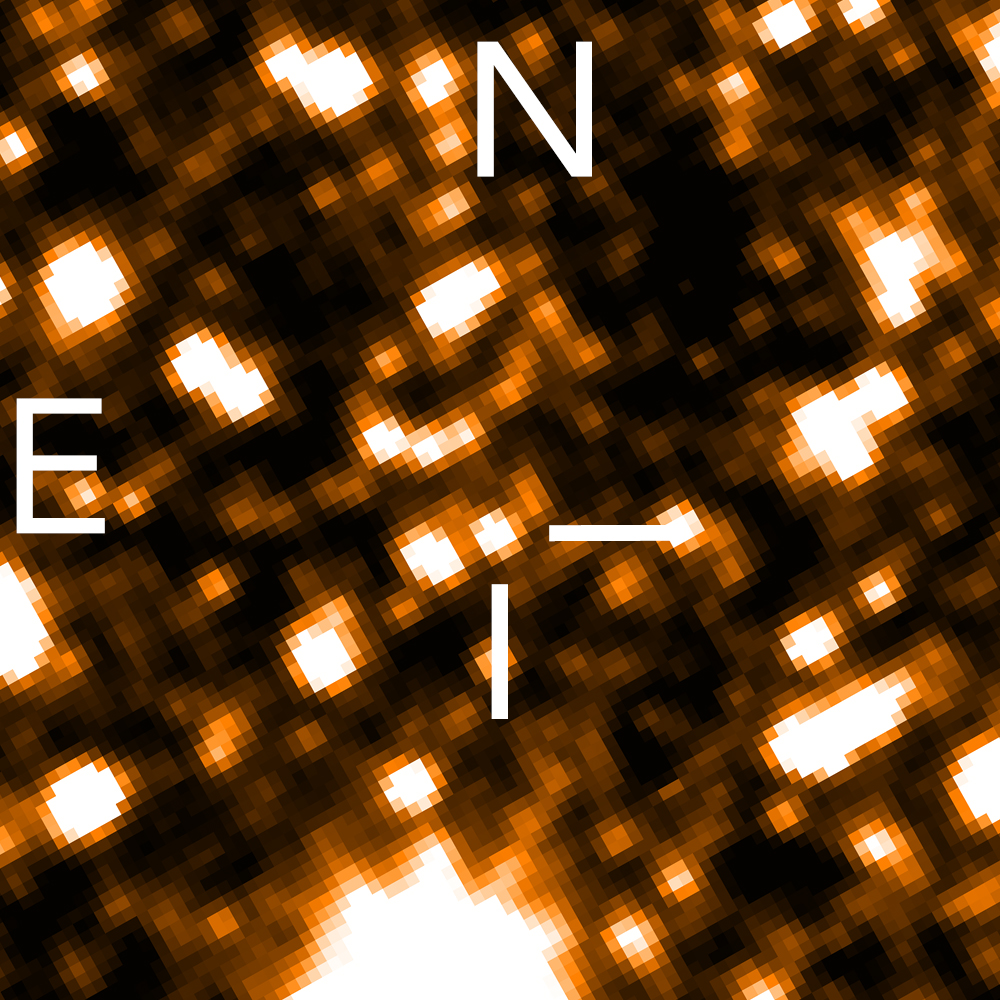}
\hfill
\includegraphics[width=0.45\columnwidth]{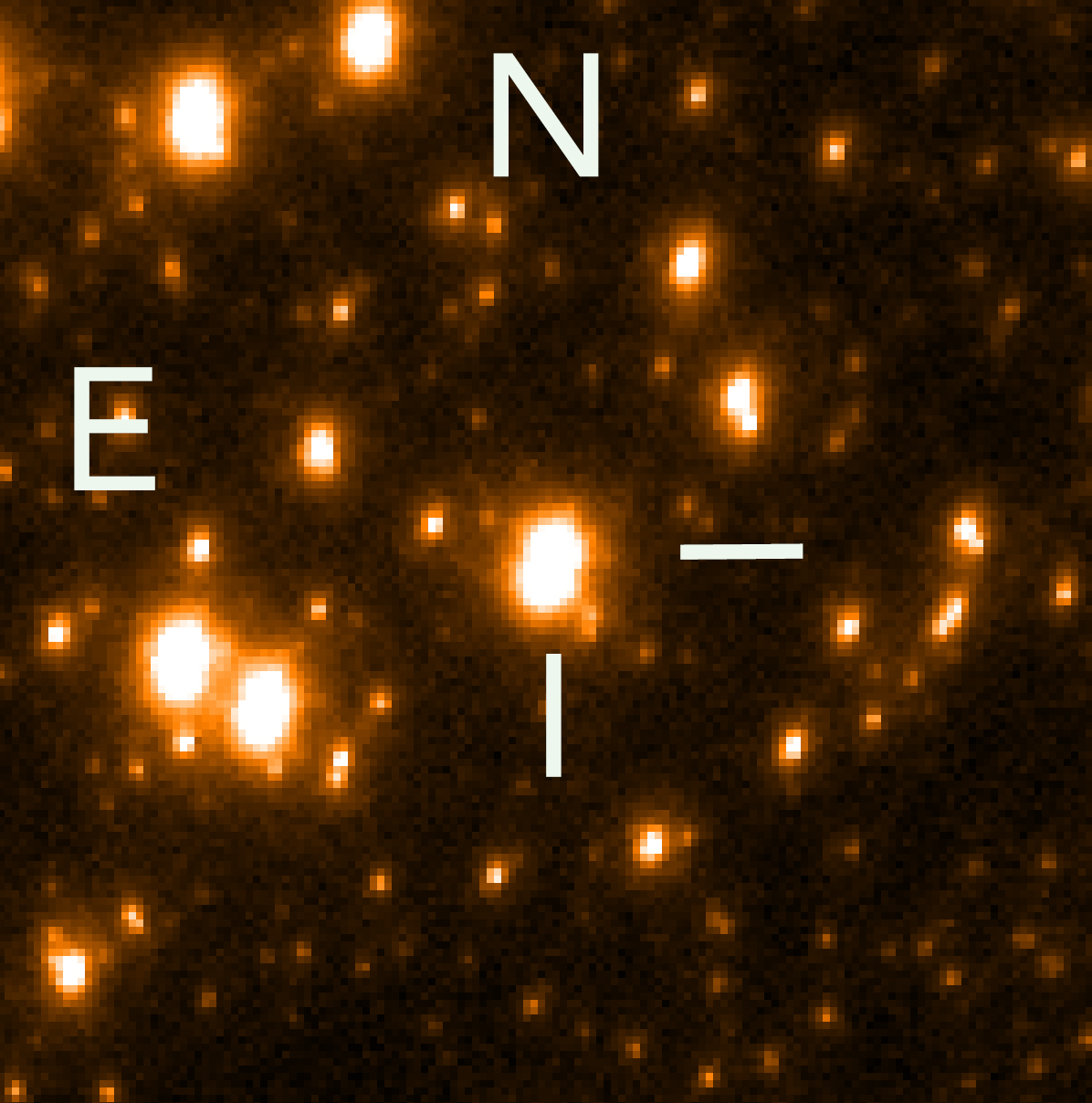}
\caption{Left panel is a 30 arcsec wide finding chart in $K$ band obtained by the VISTA 4 m telescope. The right panel is a 6 arcsec zoom obtained by Keck II. {The elongation of the point spread function of the target and of the field star is clearly visible. This is due to the observations conditions rather than to a companion.}}
\label{fig:keck}
\end{figure}

\begin{figure}[ht]
\includegraphics[width=0.45\textwidth]{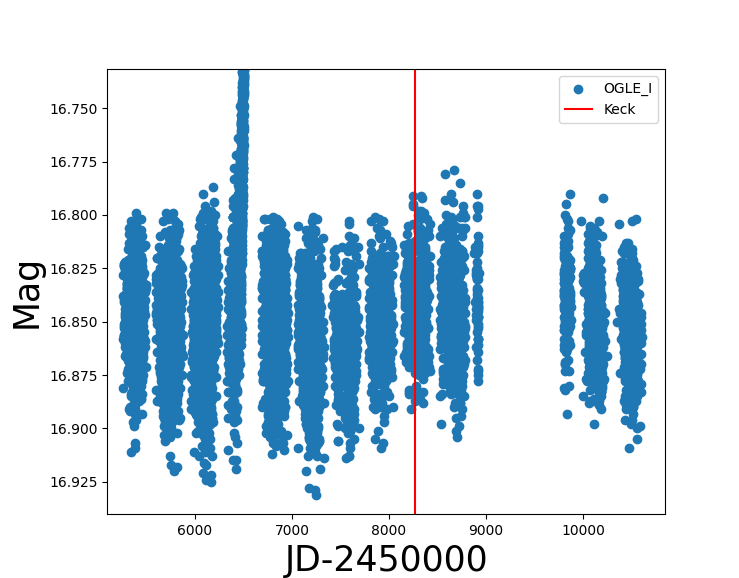}

\caption{{Zoom around the baseline of OGLE-2013-BLG-1125. A bump lasting several years is visible at the time when the Keck data were collected (indicated by the red vertical line at JD$\approx 2458264$)}.}
\label{fig:OGLEBaseline}

\end{figure}

We reprocess a 6 arcmin field of VVV \cite{Minnit2010} images obtained with the ESO Vista 4m telescope centered on our target using our standard approach {(see Beaulieu et al. 2018)}. It is calibrated against the 2MASS system \citep{Skrutskie2006}, using \textit{Gaia} as an astrometric reference. We then cross-identify the sources between VVV and Keck. We used 45 stars {(fairly bright and isolated) to calibrate to 2 \% the Keck data}. As can be seen in Fig.~\ref{fig:keck}, we {found no} clear separation between the lens and the source. Instead, we obtained their joint magnitude {$K_{S, {\rm tot}} = 14.1 \pm 0.03$} mag.
We note that at the time of Keck observations (${\rm JD}\sim2458264$), the baseline {magnitude is brighter by $\ge0.01$} mag, as can be seen in the {Fig.~\ref{fig:OGLEBaseline}}. Previous microlensing studies \citep{Wyrzykowski2006,Bachelet2022} already reported similar level of variability for red giant sources over long timescales. While the bump seen in the lightcurve was not included during the modeling, we {verified that it} did not affect the final modeling results. However, it might explain the slight calibration offset seen between VVV and Keck data.

\begin{table*}[]
\caption{Summary of data used in this analysis.}
\label{table:sites}
\centering 
\begin{tabular}{l c c c c  c c l }
\hline\hline
 Name & $N_{\mathrm{data}}$ & $N_{\mathrm{model}}$ & Aperture [m]  & Location &  Filter & $\Gamma_\lambda$ & Pipeline \\

 \hline
$\rm{OGLE_{I}}$ & 17054 & 4609 & 1.0 & Chile & $I$ & 0.41 & {\tt OGLE} \\
$\rm{OGLE_{V}}$ & 104 & 44 & 1.0 & Chile & $V$ & 0.62 & {\tt OGLE} \\
$\rm{MOA_{R}}$ & 18078 & 6040 & 1.8 & New Zealand & $R$ & 0.51 & {\tt Bond} \\
$\rm{MOA_{V}}$ & 286 & 0 & 1.8 & New Zealand & $V$ & 0.62 & {\tt Bond} \\

$\rm{VVV_K}$ & 88 & 37 & 4.1 & Chile & $K_s$ & 0.22 & {\tt Beaulieu} \\
$\rm{\mu FUN_I}$ & 15 & 15 & 1.3 & Chile & $I$ & 0.41 & \\
$\rm{\mu FUN_V}$ & 14 & 14 & 1.3 & Chile & $V$ & 0.62 & \\
$\rm{COJA_{ip}}$ & 61 & 61 & 1.0 & Australia & $i'$ & 0.48 & {\tt DanDIA} \\
$\rm{CPTA_{ip}}$ & 317 & 317 & 1.0 & South Africa & $i'$ & 0.48 & {\tt DanDIA} \\
$\rm{CPTB_{ip}}$ & 13 & 13 & 1.0 & South Africa & $i'$ & 0.48 & {\tt DanDIA} \\
$\rm{LSCA_{ip}}$ & 454 & 454 & 1.0 & Chile & $i'$ & 0.48 & {\tt DanDIA} \\
$\rm{LSCB_{ip}}$ & 36 & 36 & 1.0 & Chile & $i'$ & 0.48 & {\tt DanDIA} \\
$\rm{Danish_Z}$ & 25 & 25 & 1.0 & Chile & $Z$ & 0.48 & {\tt DanDIA} \\

\hline
$\rm{Total}$ & 36545 & 11665 && \\
\hline
\end{tabular}
\tablefoot{$N_{\rm data}$ is the total number of observations while $N_{\rm model}$ indicates the number of data used during the modeling for each datasets. $\Gamma_\lambda$ are the limb-darkening coefficients used for the modeling.}
\end{table*}

\section{Light curve analysis}
\label{sec:analysis}
\subsection{Modeling}

{The modeling of this event began while the event was still in progress}. Indeed, real-time modeling by {\tt RTModel} \footnote{\url{http://www.fisica.unisa.it/gravitationAstrophysics/RTModel/2013/RTModel.htm}}, the modeling software described in \citep{Bozza2024}, started on the 2013 October 6. More data were requested because a planetary solution was likely. On {2013 October 13}, the presence of a parallax signal was established, based on binned online reductions of survey- and follow-up data. The event assessment thread was concluded on {October 28} with the detection of a weak orbital motion signal. {Independently, Cheongho Han contributed to the real-time assessment of the event\footnote{\url{http://astroph.chungbuk.ac.kr/~cheongho/modelling/2013/model_2013.php}} while it was still in progress}. The main purpose of the {real-time} modeling is to optimize follow-up observations. Nevertheless, all algorithms converged to a similar and unique resonant caustic model, due to a planetary lens with $q<0.005$.

\begin{figure*}[!ht]
\resizebox{\hsize}{!}{\includegraphics{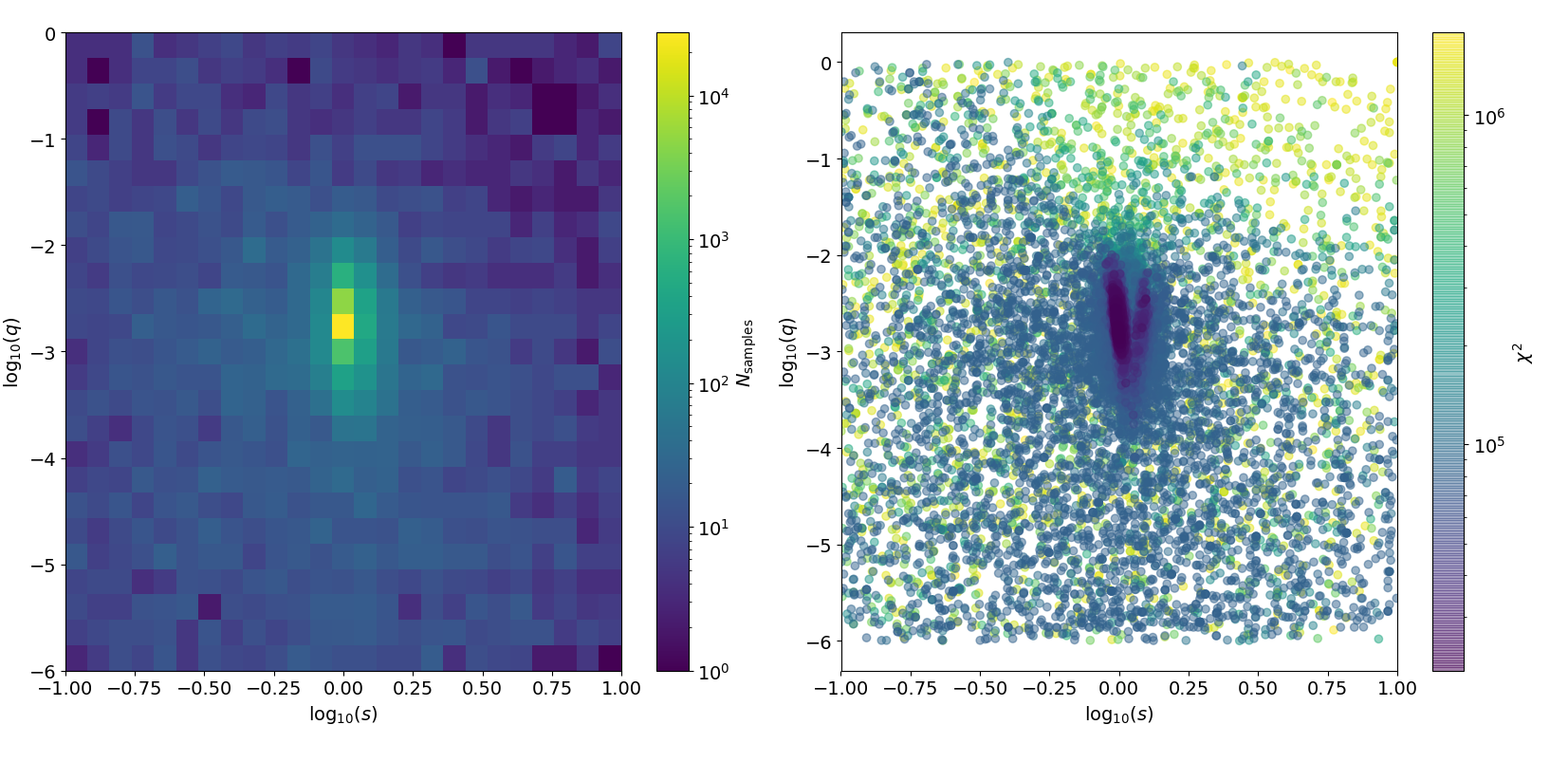}}
\caption{A map of the $\log{s}, \log{q}$  distribution of the differential evolution samples is shown as 2D histogram (left) for the static case along with the actual $\chi^2$ of each population element (right).\label{fig:des}}
\end{figure*}

For this analysis, we started with an initial grid search, the best solution in terms of mass ratio, separation and track angle agreed with those reported online by {\tt RTModel} \footnote{\url{https://projects.phys.unisa.it/gravitationAstrophysics/RTModel/2013/OB131125.htm}}. Thereafter the modeling run was restarted relying on the differential evolution method as implemented in the {\tt pyLIMA} modeling package\footnote{\url{https://ebachelet.github.io/pyLIMA/pyLIMA.html}} by \cite{bac17} and the affine-invariant ensemble sampler \citep{for13} to generate Markov Chain Monte Carlo (MCMC) chains to sample the parameters around the best fit solution. The method itself probes the whole parameter space in a heuristic way and {with uniform priors}. {Due to the larger number of parameters,} the run was made with a population size of 20. The best fit has a lens geometry with intermediate topology. That is to say, there is only a single caustic with six cusps while a close or wide configuration would make a degenerate solution more likely. The population of the differential evolution samples probing the parameter space is shown in Fig.~\ref{fig:des}. 

\begin{figure*}[!ht]
\includegraphics[width=0.9\textwidth]{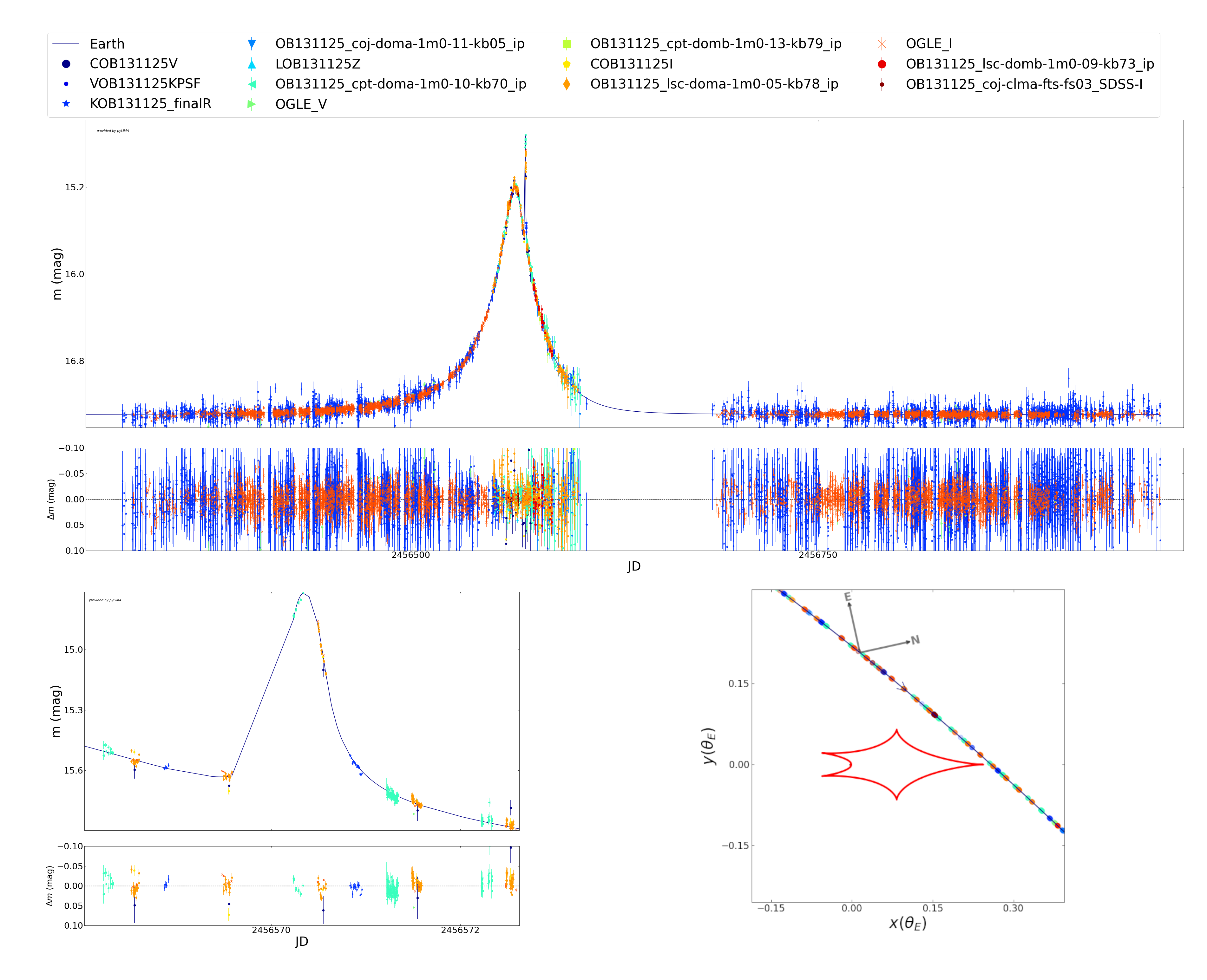}

\caption{Top: Light curves and best-fit parallax model $u_0>0$ for OGLE-2013-BLG-1125/MOA-2013-BLG-584 are shown. Bottom left: Zoom at the time of the anomaly. Bottom right: The corresponding source trajectory and resonant caustic. East and North directions are also represented at the time $t_{\rm 0,par}$.}
\label{fig:lightcurves}
\end{figure*}

{Because the anomaly visible in Fig.~\ref{fig:lightcurves} can potentially be explained by a double-source model, as described by \citet{Griest1992}, we also explored this possibility. The binary-source models are, however, strongly disfavored ($\Delta \chi^2>1000$), and we therefore did not investigate this hypothesis further}.

We then study the possibility of second-order effects. The motion of Earth around the Sun can be {imprinted} on the microlensing light curve as an annual parallax effect, particularly for events with durations comparable to or longer than the Earth's orbital period \citep{Poindexter2005}. While the timescale of OGLE-2013-BLG-1125 is $t_{\rm E}\sim35$ days, the parallax is clearly visible in the {MOA and OGLE} data. Therefore, we investigate {this} possibility by fitting $\pi_{\rm EN}$ and $\pi_{\rm EE}$, the (equatorial) North and East components of the microlensing parallax vector, and found that the models are strongly favored (with $\Delta \ln \mathcal{L}>1000$), as can be seen in Table~\ref{tab:best_fit}. As we can see in the next section, this is because the lens is relatively low-mass and close by. If $d_{\rm L}<<d_{\rm S}$, this implies that $\pi_{\rm E}\propto \frac{1}{M_{\rm L}d_{\rm L}}$, i.e. this effect is more visible for closer and lighter hosts, which is the case here. As usual with lenses located towards the Galactic Bulge, we found four degenerate models, the signature of the binary ecliptic degeneracy with $\pm u_0$ \citep{Skowron2011}. This also applies to the $\pm \pi_{\rm EE,EN}$ degeneracy which is also referred to as the jerk-parallax degeneracy \citep{gould2004}.

It is well known that the orbital motion of the lens during the course of the event can impact the modeling of the lightcurve \citep{Dominik1998}. Therefore, we investigate such possibility by modeling the orbital motion of the lens with a circular orbit, as detailed in \citep{Bozza2018}. However, we find no improvement in the model significance (with $\Delta \ln \mathcal{L}\sim0$). Similarly, the orbital motion parameters were poorly constrained and consistent with 0, while the parallax parameters almost remain unchanged. This can be explained by the details of the lensing system presented in the next section. Indeed, with $M_{\rm L}\sim0.1M_\odot$ and $d_{\rm L}\sim1.2$ kpc, one can estimate a projected separation $a_\perp\sim1$ au and ultimately a lower-limit on the planetary period of $P>3$ years (assuming a circular orbit). Therefore, the event duration represents at best 6\% of the orbital period, implying a difficult measurement.

{When finite-source effects are detected in the lightcurve}, the limb-darkening effect has to be taken into account to accurately measure $\rho$ \citep{Witt1995}. As detailed in the next section, we found the source to be a K0 giant. Assuming solar metallicity, $T_{\rm eff}=5000$ K and {\rm $\log(g)=3.5$}\,cgs, we used \citet{Claret2011} to compute the limb-darkening coefficients $\Gamma_\lambda$ presented in the Table~\ref{table:sites}. We note that we assumed $\Gamma_R=(\Gamma_V+\Gamma_I)/2$ for the non-standard MOA $R$ filter. This hypothesis is not considered to be dramatic as the MOA $R$ data set covers only the tail of the planetary anomaly, where finite-source effects are minimal. Overall, we found the limb-darkening effect to be small but we use the modeling results that include this effect for the rest of this analysis.

During the final modeling step, the exploration of degenerate minima with MCMC, pyLIMA includes an errorbar rescaling of the errorbars for all datasets in the form:
\begin{equation}
\sigma'_{\rm j}(t) = 10^{\log{k_j}} \sigma_j(t) ,
\end{equation}
where $\sigma_j(t)$ is the original error in flux unit and $\log(k_j)$ is an additional free parameter for each dataset $j$.

\begin{table*}[h!]
\centering
\caption{Best fit model parameters for different models.} \label{tab:fitparams}
\begin{tabular}{lccccc}
\hline
\hline

  Parameter &Static& $u_0> 0 ~;~ \pi_{\rm EN}> 0$ & $u_0> 0 ~;~ \pi_{\rm EN}< 0$ & $u_0< 0 ~;~ \pi_{\rm EN}> 0$ & $u_0< 0 ~;~ \pi_{\rm EN}< 0$\\
\hline
$t_0 ~[days]$ & $2456563.891\pm 0.007$&$2456564.30 \pm 0.02$ & $2456564.006 \pm 0.008$ & $2456563.942 \pm 0.009$ & $2456564.23 \pm 0.01$\\
$u_0 ~[\theta_E]$ & $0.1032\pm 0.0008$&$0.163 \pm 0.002$ & $0.170 \pm 0.002$ & $-0.168 \pm 0.002$ & $-0.165 \pm 0.002$\\
$t_{\rm E} ~[days]$ & $50.5\pm0.3$&$36.0 \pm 0.3$ & $33.8 \pm 0.3$ & $34.0 \pm 0.3$ & $35.3 \pm 0.3$\\
$\rho$ & $0.00254\pm0.00005$&$0.00438 \pm 0.0001$ & $0.0048 \pm 0.0001$ & $0.0048 \pm 0.0001$ & $0.0045 \pm 0.0001$\\
$s ~[\theta_E]$ & $0.
986\pm0.001$&$1.046 \pm 0.002$ & $1.043 \pm 0.002$ & $1.038 \pm 0.002$ & $1.047 \pm 0.002$\\
$q$ &$0.00166\pm0.00002 $&$0.00194 \pm 0.00002$ & $0.00233 \pm 0.00004$ & $0.00244 \pm 0.00004$ & $0.00202 \pm 0.00002$\\
$\alpha ~ [rad]$ & $2.465\pm0.001$&$2.369 \pm 0.003$ & $2.46 \pm 0.003$ & $3.798 \pm 0.003$ & $3.895 \pm 0.003$\\
$\pi_{\rm EN} $ & ... &$0.78 \pm 0.05$ & $-0.55 \pm 0.04$ & $0.84 \pm 0.05$ & $-0.55 \pm 0.04$\\
$\pi_{\rm EE} $ & ... &$0.70 \pm 0.02$ & $0.71 \pm 0.02$ & $0.71 \pm 0.02$ & $0.70 \pm 0.02$\\
\hline
$\chi^2$ &$30083$ &$24292 $ & $24085$ & $24171$ & $24144$\\
\hline
\end{tabular}
\tablefoot{ The reference date of $t_{ \rm 0,par} = 2456560$. The $\chi^2$ values are computed at each solution without uncertainties rescaling.}
\label{tab:best_fit}

\end{table*}

\subsection{Source Properties}
\label{sec:sourceproperties}
\begin{figure}[!ht]
\includegraphics[width=0.45\textwidth]{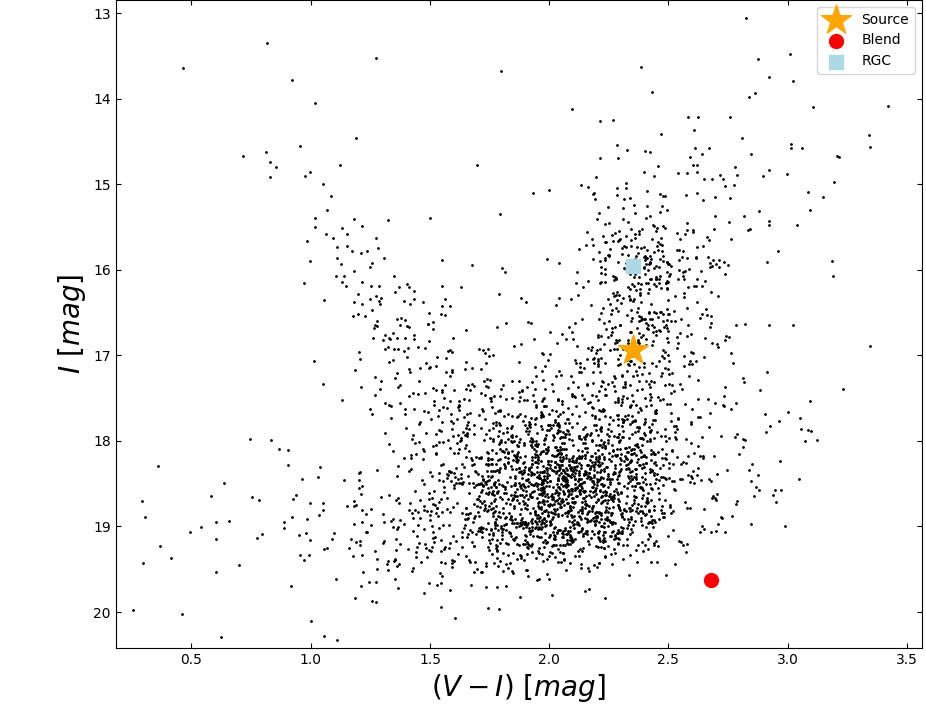}

\caption{OGLE-IV color-magnitude diagram. The source is most likely a K0 red giant located in the Galactic Bulge.}
\label{fig:CMD}
\end{figure}

From the models, we obtain an apparent magnitude and color of the source $[I,(V-I)]_S=[16.95\pm0.01,2.35\pm0.01]$ mag from the OGLE {color magnitude diagram}. As can be seen in Fig.~\ref{fig:CMD}, the source is a K0 giant. Indeed, assuming the apparent magnitude and color of the red giant clump is $[I,(V-I)]_{\rm 0,RGC}=[14.32, 1.06]$ mag, one can derive that $[A_I,E(V-I)]=[1.69\pm0.01,1.294\pm0.007]$ mag, in good agreement with OGLE website\footnote{\url{https://ogle.astrouw.edu.pl/}} values of $[1.56,1.27]$ mag. Using the average leads to a color and magnitude of the source  $[I,(V-I)]_{\rm 0,S}=[15.3\pm0.1,1.1\pm0.1$] mag. We note that the infrared source magnitude from the VVV dataset is $K_S=14.30$ mag, leading to $(V-K_s)=4.99$ mag and $(V-K_s)_{\rm 0,S}=2.24$ mag,  in excellent agreement with our $(V-I)_{\rm 0,S}$ measurement for a K0 giant \citep{Bessell1988}. This is in excellent agreement with the measurements of \citet{Bensby2017}.
Using \citet{Adams2018}, we found the angular source radius $\theta_*=4.1 \pm 0.4 \rm \ \mu as$ and  ultimately we derive $\theta_{\rm E}=\frac{\theta_*}{\rho}=0.85\pm0.08$ mas and $\mu_{\rm rel} = \frac{\theta_{\rm E}}{t_{\rm E}}=9.2\pm 1.0 \ {\rm mas \ .yr^{-1}}$.

\section{Discussion and conclusion}
\label{sec:interpretation}

From the lightcurve modeling, we obtain two constraints on the mass and distance of the lens with the measurements of $\theta_{\rm E}$ and $\boldsymbol{\pi}_E$. Eq.~(\ref{eq:mldl}) immediately leads to $M_{\rm L}\approx0.1 \ M_\odot$ and $d_{\rm L}\approx1.2$ kpc if we assume $d_{\rm S}=8$ kpc. 
But we obtain a third constraint from the measurement of the baseline flux in $K_{s,{\rm tot}}$ from the Keck images. This offers an opportunity to check for consistency between our results. Using the previous results, the magnitude of the lens is predicted to be $K_{\rm S,{\rm L}}\sim19.5$ mag with a separation $\delta \approx \mu_{\rm rel}\times5\approx50$ mas {(the timespan between the event peak and Keck observations is roughly 5 years)}. This makes the detection of the lens challenging in the Keck images, as we already discussed in the Sect.~\ref{sec:KECK}.

\begin{figure}[!ht]
\includegraphics[width=0.45\textwidth]{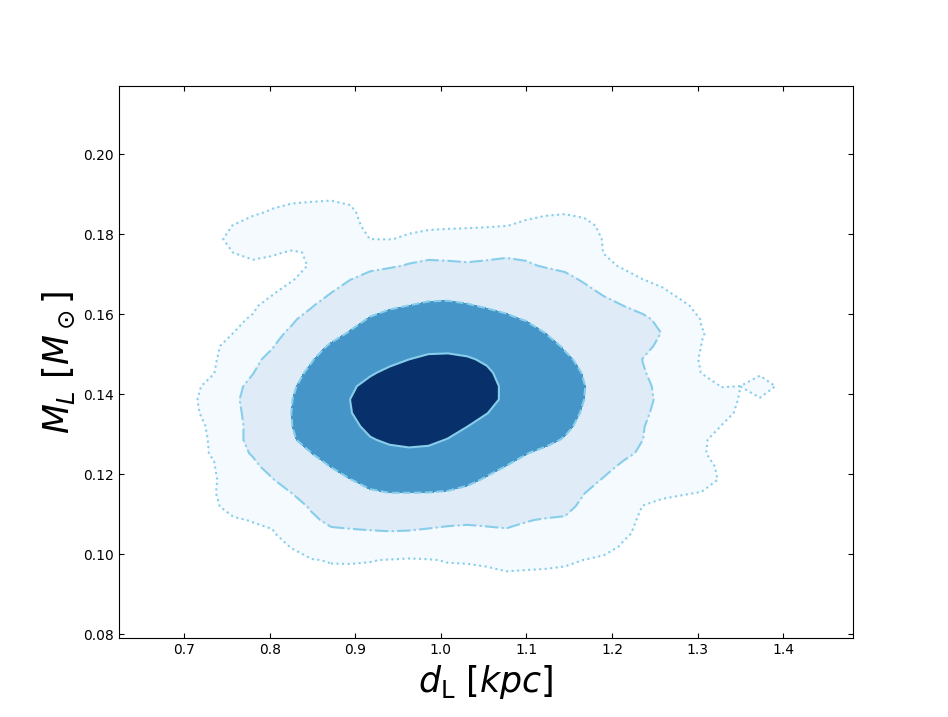}
\caption{{\tt pyLIMASS} distribution for $M_{\rm L}$ vs $d_{\rm L}$. $39\%, 86\%,99\%$ and $100\%$ contours are presented from dark to light blue.}
\label{fig:pyLIMASS}
\end{figure}

We can still combine all the constraints in a Bayesian framework with the {\tt pyLIMASS} software \citep{Bachelet2024}. Using $V,I,K,\rho,t_{\rm E},\pi_{\rm EN},\pi_{\rm EE}$ observables from the best model ($u_0\ge0 ; \pi_{\rm EN}\leq 0$) , the Keck baseline measurement $K_{\rm s,{\rm tot}}$, the estimated visual absorption $A_V,$ and the stellar parameters from \citet{Bensby2017}, {\tt pyLIMASS} finds no tensions within these observables and estimates the parameters presented in the Table~\ref{tab:physparams} and Fig.~\ref{fig:pyLIMASS}. The lens is confirmed to be a late-type M dwarf with a mass $M_{\rm L}=0.139\pm0.011 \ M_\odot$ located at $d_{\rm L}=1.0\pm0.1$ kpc from the Earth. Consequently, the companion is a Saturn-mass planet with $M_{\rm p}=105\pm13  \ M_\oplus$ located beyond the snowline of the system, i.e. $a\ge a_\perp\sim1$ au.

\begin{table}[!ht]
\centering
\caption{Best fit physical parameters and derived estimates from {\tt pyLIMASS}} \label{tab:physparams}
\begin{tabular}{l c}
\hline\hline
Parameter & Estimates \\
\hline
$d_{\rm S}\left(\mathrm{kpc}\right)$& $6.2\pm2.3$ \\

$\theta_{\rm \star}\left( \mathrm{\mu as} \right)$ & $4.35\pm0.29$ \\
$d_{\rm L}\,\left(\mathrm{kpc}\right)$& $1.000\pm0.082$ \\
$M_{\rm L}\,\left( M_{\odot} \right)$ & $0.139\pm0.011$ \\
$M_{\rm p}\,\left( M_{\oplus} \right)$ & $105\pm13$ \\
$\theta_{\rm \mathrm{E}} \left( \mathrm{mas} \right)$ & $0.976\pm0.056$ \\
$\mu_{\rm \mathrm{rel}} \left( \mathrm{mas}/\mathrm{yr} \right)$ & $10.61\pm0.67$ \\
\hline
\end{tabular}

\end{table}

\begin{figure*}
\resizebox{\hsize}{!}{\includegraphics{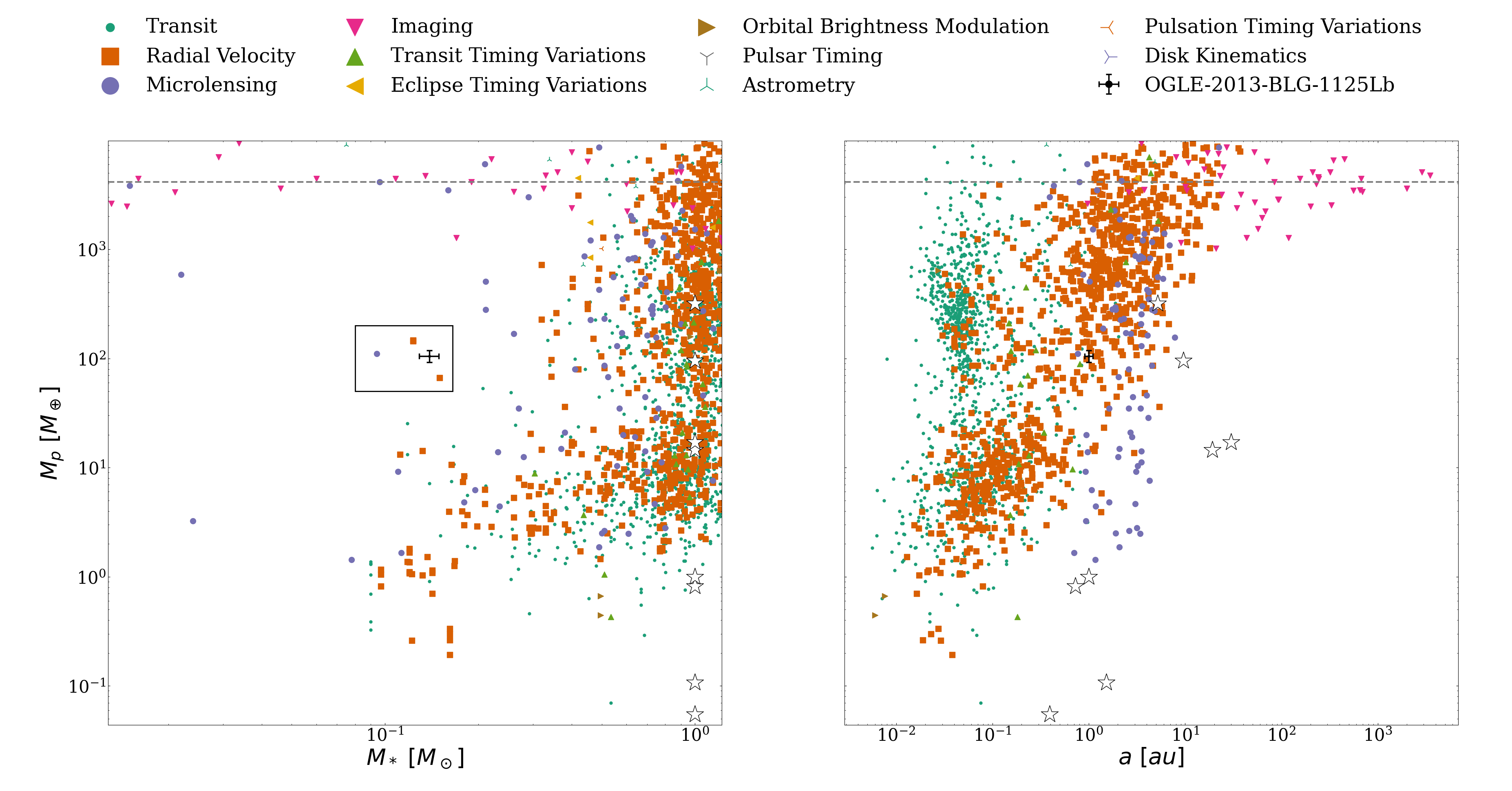}}%
\caption{
{Confirmed planets as listed by NASA Exoplanet Archive (accessed on 2026 July 3) with host mass measured at better than 50 $\%$. The black symbol with errorbars represents the masses estimate of OGLE-2013-BLG-1125/MOA-2013-BLG-584Lb from this study. The gray dash line indicates the deuterium burning mass limit (defined as 0.013 $M_\odot$). The empty star symbols represent the planets of the Solar System  Left: Planet mass (or $m_p\sin i$ for radial velocity planets) versus host mass. The rectangle highlights five planets discussed in Section~\ref{sec:interpretation}. Right: Planet mass versus planet semi-major axis. For the majority of planets discovered by microlensing, including OGLE-2013-BLG-1125Lb, the projected separation $a_\perp$ (i.e. a lower limit on the true separation $a$) is plotted. When several entries of the same planets exist, the default is plotted (defined as \textquotesingle 1\textquotesingle{} by the NASA Exoplanet Archive
).}}
\label{fig:planets}
\end{figure*}

 As can be seen in Fig.~\ref{fig:planets}, with a {Saturn-mass} located at $a_\perp\sim1$ au form its host, OGLE-2013-BLG-1125Lb is a typical microlensing detection of a cold planet beyond the snowline. Indeed, microlensing is capable of detecting a large variety of planets located beyond the snowline \citep{Suzuki2018,Mroz2023,Han2024a,Han2024b,Yee2025,Han2026a,Han2026b}. These systems are difficult to study with other methods of detection, due to intrinsic observational {limitations}. But they are of special interest as they provide unique constraints on planet formation models \citep{Mercer2020,Pinilla2022,Helled2026}. For instance, it seems almost impossible to form giant planets for hosts with $M_*\le0.5M_\odot$ via core accretion \citep{Burn2021,Emsenhuber2021}. This {is understood to be due to} the higher efficiency of radial drift \citep{Pinilla2022,Burn2024}, which limits the growth of planetesimals and ultimately reduces the giant-planet formation efficiency around low-mass stars. Similarly, the gravitational-instability mechanism generally requires massive disks to form giant planets \citep{Mercer2020,Pinilla2022,Helled2026}. Moreover, planets formed via disk fragmentation are expected to originate at large orbital separations (typically $\gtrsim 50$ au), which is significantly wider than the separations probed by the current microlensing population ($\lesssim 10$ au).

 However, giants planets around low-mass stars {exist} and are also detected in transit \citep{Bryant2023,Dransfield2026} and radial velocity \citep{Marcy2001,Trifonov2018,Morales2019,Quirrenbach2022} surveys. Although located in an empty part of the mass-mass diagram presented in Fig.~\ref{fig:planets}, OGLE-2013-BLG-1125Lb is surrounded by four other planets as listed in Table~\ref{tab:4planets}. Detected via radial velocity, GJ 3512 b,c \citep{Morales2019,Morales2025} and GJ 9066 c \citep{Quirrenbach2022} are Saturn-mass planets at $\sim 1$ au of $0.123\pm0.005 \ M_\odot$ and $0.150\pm0.010 \ M_\odot$ hosts, respectively. {KMT-2018-BLG-1190Lb \citep{Ryu2019} is comparable to OGLE-2013-BLG-1125Lb, but its physical parameters are less precisely constrained, primarily due to the less precise measurement of the microlensing parallax $\pi_E$ and the absence of high-resolution imaging confirmation at the time of writing. With a sample of five}, it is not clear yet if this population is significant or not but  
additional or alternative formation pathways {seem to be} required to explain this population, such as enhanced pebble accretion models \citep{Dash2020,Liu2020,Qiao2023,Pan2024} or a new migration process \citep{Sanchez2025}.

The upcoming Roman Galactic Bulge Survey \citep{Spergel2015,Akeson2013,Penny2019}, which is expected to find several thousands of cold planets via microlensing and transit \citep{Wilson2023,Edmondson2026}, is especially well designed to explore this part of the parameter space and, ultimately, test a wide variety of planet formation mechanisms with unprecedent precision.

\begin{table*}
\centering
\renewcommand{\arraystretch}{1.3}
\caption{Known Saturn-mass planets around low-mass hosts.} \label{tab:4planets}
\begin{tabular}{l lcccc}
\hline\hline
Planet name & Reference & Detection & Host mass & Planet mass & Semi-major axis \\
\hline

GJ 3512 b & \cite{Morales2025} & Radial velocity & $0.123\pm0.005 M_\odot$ & $247^{+13}_{-86} M_\oplus$ & 0.34 au \\
GJ 3512 c & \cite{Morales2025} & Radial velocity & $0.123\pm0.005 M_\odot$ & $245^{+13}_{-83} M_\oplus$ & 1.7 au \\
GJ 9006 c & \cite{Quirrenbach2022} & Radial velocity & $0.150\pm0.010 M_\odot$ &$ \le 68 M_\oplus$ & 0.9 au  \\

KMT-2018-BLG-1990Lb & \cite{Ryu2019} & Microlensing & $0.141^{+0.202}_{-0.061} M_\odot$&$180^{+250}_{-80} M_\oplus$& $\ge 1.039^{+0.951}_{-0.351}$ au\\

OGLE-2013-BLG-1125Lb  & This work & Microlensing & $0.139\pm0.011 M_\odot$ & $105\pm13 M_\oplus$ & $\ge 1.0$ au \\

\hline

\hline
\end{tabular}

\end{table*}

\begin{acknowledgements}
     {The authors thank the anonymous reviewer for helpful comments that significantly improved the manuscript.}
     Work by R.A.S and EB was supported by NASA grant NNX15AC97G.  The work  by  CR  was supported  by  an  appointment  to the NASA Postdoctoral Program  at the  Goddard  Space  Flight  Center, administered by USRA through a contract with NASA. VB was supported by PRIN 2022 CUP D53D23002590006. This work made use of observations from the LCOGT network, which includes three SUPAscopes owned by the University of St. Andrews. The MOA project is supported by JSPS KAKENHI Grant Number JSPS24253004, JSPS26247023, JSPS23340064, JSPS15H00781, and JP16H06287. This work has made use of data from the European Space Agency (ESA) mission {\it Gaia} (\url{https://www.cosmos.esa.int/gaia}), processed by the {\it Gaia} Data Processing and Analysis Consortium (DPAC, \url{https://www.cosmos.esa.int/web/gaia/dpac/consortium}). This research has made use of the NASA Exoplanet Archive, which is operated by the California Institute of Technology, under contract with the National Aeronautics and Space Administration under the Exoplanet Exploration Program. Based on data products from VVV Survey observations made with the VISTA telescope at the ESO Paranal Observatory under programme ID 179.B-2002. J.C.Y. acknowledges support from U.S. NSF Grant No. AST-2108414 and U.S. NASA Grant No. 80NSSC25K7146. Data presented herein were obtained at the W. M. Keck Observatory from telescope time allocated
477 to the National Aeronautics and Space Administration through the agency’s scientific partnership
478 with the California Institute of Technology and the University of California. The Observatory was
479 made possible by the generous financial support of the W. M. Keck Foundation.
The material presented here is also based upon work supported by NASA under award number 80GSFC21M0002. This work was supported by the University of Tasmania through the UTAS Foundation and the endowed Warren Chair in Astronomy and the ANR SPACE-MLENS. This research was also
486 supported in part by the Australian Government through the Australian Research Council Discovery
487 Program grant awarded to Cole and Beaulieu. 
The OGLE project has received funding from the Polish National Science
Centre grant OPUS-28 2024/55/B/ST9/00447 awarded to AU.The MOA project is supported by JSPS KAKENHI Grant Number  

JP16H06287, JP22H00153, JP23KK0060 and JP25H00668. This work was supported by the University of Tasmania through the endowed Warren Chair in Astronomy. J-PB, JB  have been supported by the Australian Government through the Australian Research Council Discovery Project Grants 200101909 and 240101842.  Research in France has been supported through the SPACE-MLENS ANR grant ANR-24-CE31-3263. GD acknowledges support by UKRI-STFC grants: ST/T003081/1 and ST/X001857/1

\end{acknowledgements}


\bibliographystyle{aa}
\bibliography{microlensing.bib}

\end{document}